\documentclass[a4paper,fleqn]{cas-dc}

\usepackage[numbers]{natbib}
\usepackage[ruled,vlined]{algorithm2e}
\usepackage{hyperref}
\usepackage[normalem]{ulem}
\usepackage{xcolor}
\usepackage{enumitem}

\def\tsc#1{\csdef{#1}{\textsc{\lowercase{#1}}\xspace}}
\tsc{WGM}
\tsc{QE}
\tsc{EP}
\tsc{PMS}
\tsc{BEC}
\tsc{DE}

\begin{document}
\let\WriteBookmarks\relax
\def\floatpagepagefraction{1}
\def\textpagefraction{.001}
\shorttitle{Out of distribution detection}
\shortauthors{Zaid et~al.}

\title [mode = title]{Out of distribution detection for skin and malaria images}

\author[1]{Muhammad Zaid}
\address[1]{Intelligent Machine Lab, Information Technology University, Pakistan}
\author[1]{Shafaqat Ali}
\author[1]{Mohsen Ali}
\author[2]{Sarfaraz Hussein}
\author[3]{Asma Saadia}
\author[1]{Waqas Sultani}

\ead{waqas.sultani@itu.edu.pk}




\address[2]{Machine Learning and Data Science @ The Home Depot, USA}
\address[3]{Central Park Medical College, Lahore, Pakistan}


\begin{abstract}
Deep neural networks have shown promising results in disease detection and classification using medical image data. However, they still suffer from the challenges of handling real-world scenarios especially reliably detecting out-of-distribution (OoD) samples. We propose an approach to robustly classify OoD samples in skin and malaria images without the need to access labeled OoD samples during training. Specifically, we use metric learning along with logistic regression to force the deep networks to learn much rich class representative features. To guide the learning process against the OoD examples, we generate ID similar-looking examples by either removing class-specific salient regions in the image or permuting image parts and distancing them away from in-distribution samples. During inference time, the K-reciprocal nearest neighbor is employed to detect out-of-distribution samples. For skin cancer OoD detection, we employ two standard benchmark skin cancer ISIC datasets as ID, and six different datasets with varying difficulty levels were taken as out of distribution. For malaria OoD detection, we use the BBBC041 malaria dataset as ID and five different challenging datasets as out of distribution. We achieved state-of-the-art results, improving 5\% and 4\% in TNR$@$ TPR95\% over the previous state-of-the-art for skin cancer and malaria OoD detection respectively.

\end{abstract}

\begin{keywords}
Skin cancer, 
\sep Malaria,
\sep Out of distribution, 
\sep Unsupervised approach,
\sep Tuplet loss, 
\sep K-reciprocal neighbors
\end{keywords}

\maketitle

\section{Introduction}

Recent years have witnessed tremendous success in applying deep neural networks to diagnose and analyze several diseases including skin cancer 
 \citep*{JBI-Skin, esteva2017dermatologist,yu2018melanoma} 
 and malaria \citep*{JBI-Malaria, LEE2021104151,SANTOSH2020103859}.
 Skin cancer is one of the most common types of cancer. There are around 1.19 million new cases of skin cancer only in 2020 \citep{sung2021global}.
 An early diagnosis, including the detection of cancer and its correct classification, has been correlated to a high rate of overall survival. 
 Similarly, according to the World Health Organization, only in 2019, 229 million cases of malaria occurred worldwide with 409,000 deaths globally due to malaria \cite{world2019world}.
 Therefore, it is very important to develop a computer-aided diagnosis system where computational methods can be used to assist medical practitioners in the early detection of different diseases including skin cancer and malaria.

 One of the key advantages of deep learning models is their  {capability to generalize}
  i.e., the ability to perform well on the testing data with several variations.  {These} variations occur due to the different image capturing sensors, lighting conditions, and resolutions. Although  {having} remarkable performance,  deep neural networks exhibit overconfident incorrect predictions
on images that are outside the training distribution. Experiments have shown that adding small noise to an image causes the network to incorrectly predict with high probability \cite{hendrycks2016baseline}. Similarly, as shown in \cite{pacheco2020out},  the image containing a dog can wrongly be classified as a skin cancer image using recent CNN architectures such as DenseNet, MobileNet, ResNet, and VGGNet, etc. This may result in
catastrophic failures in real-world applications, especially in healthcare systems. For example, a network that is trained to detect a predefined number of classes is forced to classify a new type of disease into one of the predefined classes. Therefore, for reliable diagnosis and treatment of diseases through images, deep neural networks must be able to avoid such wrong overconfident predictions for such samples.
{Secure and reliable deployment of a healthcare system demands that the model should be accurate and vigorous to distribution change.}
This drives the necessity for methods that can efficiently detect out-of-distribution (OoD) samples.

 \begin{figure*}
\includegraphics[page={4},width = \linewidth]{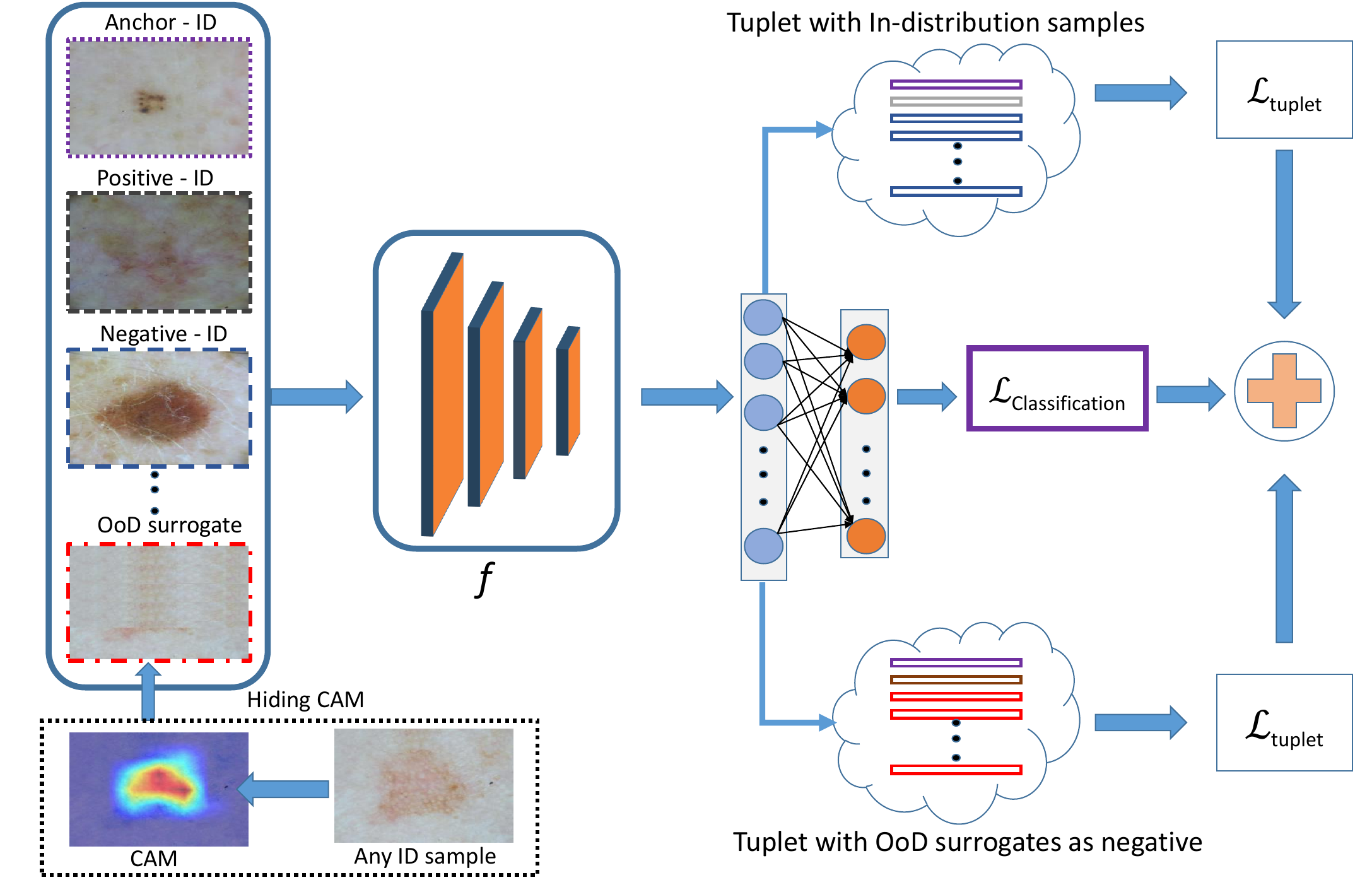}
    \caption{The flow diagram of the proposed OoD detection approach for skin cancer images. Before passing the images through CNN network $f$, we generate OoD surrogate using `HideCam'. Two types of tuplet are being created afterwards, i.e. one contains only ID (In-distribution) samples and the other contains both ID samples and OoD (Out-of-distribution) surrogates. Tuplet loss is applied independently on each tuplet and classification loss is applied on all ID samples. Exactly the same approach is used for OoD detection in malaria except that for malaria, we generate OoD surrogate using permutation. Best viewed in color.} 
    \label{fig:system}
\end{figure*}

 The distribution that generates the training data is known as In-Distribution (ID), while the ones that do not belong to training distribution are called Out-of-Distribution (OoD) samples.
 Enabling deep networks to reliably raise a flag if the sample during test time is coming from a different distribution other than training distribution is important to detect new types of diseases.
 Attempts have been made to detect OoD samples using deep learning \citep*{hendrycks2016baseline,liang2018enhancing,hsu2020generalized, sastry2020detecting} but, despite being extremely useful, little progress has been made in the medical domain \cite{JBI-OOD}. During model deployment, OoD samples occur due to images from unrelated tasks, wrongly acquired images, and the appearance of a new class of disease \cite{cao2020benchmark}. \
 Many recent methods \cite{liang2018enhancing,lee2018simple} develop supervised OoD approaches in which they assume the availability of manually annotated OoD samples during the training.
 However, manually annotated OoD samples are hard to collect and difficult to annotate especially when dealing with different medical disease images. 
 Most importantly, since any example that is not part of ID can be considered OoD, therefore any finite sampled data will not be a complete representation of the OoD. 
 
In this paper, we propose an approach to robustly classify OoD samples in skin and malaria images without needing to access labeled OoD samples during training.  
The proposed approach fine-tunes the existing classification network for learning representation such that intra-class variations are minimized. 
To discriminatively train the network and address the limitation of unavailability of OoD data during the training, we introduce OoD surrogate examples employing distribution examples. Chen et al., \cite{chen2020simple} has extensively explored different families of augmentation and reported that some augmentation techniques lead to learning better representation and some can degrade the performance as well. Furthermore, authors in \cite{tack2020csi} suggested that such performance degrading augmentations (e.g., rotation), can be used for OoD detection by considering them as negative.  Inspired by \cite{tack2020csi} and noting the fact that skin and malaria images contain different visual information, we propose two approaches to generate OoD surrogates.
For skin OoD detection, a novel class saliency-based OoD surrogate generation method, \textit{HideCam}, is designed, to create OoD datasets from ID samples. 
Our strategy results in the OoD images whose non-salient regions are similar to the non-salient regions of ID images and this could be assumed to be on the boundary of the skin classes in the feature space. Similarly, for malaria OoD detection, OoD surrogates are generated using permuting different parts of an image. 
After that, a tuplet loss-based metric learning is employed to achieve two goals. 
First to bring ID samples close to each in the feature space. Second to map such OoD images away from the class boundaries of ID images. This tight mapping of the ID samples over to the latent space, allows us to employ a retrieval strategy for identifying OoD samples. Furthermore, to maintain the discriminative power of the classifier, classification loss is employed.  Finally,  instead of thresholding distance to the naive nearest neighbor, inspired by K-reciprocal nearest neighbors \citep{zhong2017re},  we introduce a new robust OoD samples ranking strategy. We have compared the proposed approach with five state-of-the-art approaches on six OoD datasets \citep{pacheco2020out} for skin cancer and five OoD datasets for malaria detection and have obtained promising results on all evaluation metrics. In summary, the proposed approach has the following contributions:

\begin{itemize}
    \item The proposed approach neither requires the network to be trained from scratch nor does it has any specific hyperparameters to tune on the labeled OoD set.
    \item We have employed novel ways to generate OoD surrogates from ID samples.
     \item The method does not require access to any real OoD data during training.
     \item We have used a unique way to improve the OoD score using K-reciprocal neighbors.
    \item Our thorough analysis reveals that the proposed approach has encouraging results when compared with the several competitive baselines.
\end{itemize}

The organization of the paper is as follows: Section 2 describes related work, section 3 contains our proposed methodology, section 4 provides details about datasets and experiments and section 5 concludes the paper.

\section{Related work}
Several methods have been developed recently for automatic skin and malaria detection using deep learning and computer vision.
Yu et al., \citep{yu2018melanoma} and Codella et al., \citep{codella2017deep} have shown that combining deep features with local hand-crafted descriptors yields better skin classification results. Similarly, Codella et al., \citep{codella2017deep} and Gessert et al., \citep{gessert2020skin} demonstrated the use of an ensemble of many networks for lesion classification. The recent public release of International Skin Imaging Collaboration (ISIC)  archive \citep{codella2019skin,tschandl2018ham10000,codella2018skin} further facilitates 
research in skin cancer classification. {Similarly, there is a huge need for improvement for worldwide deadly malaria disease detection. Although efforts have been made by researchers using different techniques \citep*{JBI-Malaria2,lee2016application,davis2019genetic},  malaria detection is still a challenging task because of mosquito ecology and disease transmission cycle \cite{zinszer2015forecasting}.}

One of the straightforward ways of detecting OoD samples is to utilize the difference of prediction probability of OoD and ID examples \citep{hendrycks2016baseline}. This method is considered to be treated as a baseline approach for all other methods. The core idea is the hypothesis that ID samples will have a higher maximum softmax probability than OoD samples.  Liang et al., \citep{liang2018enhancing} introduced a supervised OoD detection approach (ODIN) that widens the gap between maximum softmax scores of ID and OoD samples by fine-tuning the hyperparameters on the OoD datasets. Hsu et al., \citep{hsu2020generalized} further improved ODIN by making it independent of OoD samples while using a probabilistic perspective of decomposition of confidence of predicted class probabilities. Similarly, Lee et al., \citep{lee2018simple} proposed another supervised approach which models pre-trained features on every layer as a class conditional Gaussian. A sample on the test time is then evaluated based on its Mahalanobis distance with class conditional means and variances. Similar to \citep{liang2018enhancing}, this approach uses input pre-processing and requires OoD samples to tune hyperparameters. In several recent works, this approach is named `Mahalanobis'. Similarly, Uwimana et al., \cite{uwimana2021out} used Mahalanobis distance-based confidence score for OoD sample detection in classifying malaria cells.

Hendrycks et al., \citep{hendrycks2018deep} demonstrated that an auxiliary OoD dataset can be used to discriminate between ID and OoD samples. Similarly, Tacket al., \citep{tack2020csi} interestingly proposed that certain augmentations of the same image could be considered as borderline cases of OoD samples, and treating them differently from original images should lead to more robust features for the OoD detection. 
 Recently, Sastry et al., \citep{sastry2020detecting} posed an interesting observation that joint patterns of activation and class labels assigned at the output layer should lead us to recognize class level patterns. Any sample that followed other than the already defined pattern for a class should then be considered as an OoD sample. They found this pattern using different orders of gram matrices. Pachecho et al., \citep{pacheco2020out} builds upon Gram OoD \citep{sastry2020detecting} and detected OoD samples for skin cancer. In their work,  layer-wise deviation of a sample's feature from its distribution is considered to be an indication of OoD and they have shown reasonable results  {on skin cancer-related OoD datasets}.

 Due to the recent use of self-supervised learning along with contrastive loss in several applications,  attempts have been made to use contrastive learning to detect OoD samples. Winkens et al., \citep{winkens2020contrastive} uses a famous method SIMCLR \citep{chen2020simple}  {for contrastive training}. It treated different augmentations as positive and every other image as negative to learn semantically rich features. Once the network is trained, they fit {Gaussian distributions to the activations on the training data }  
to every layers’ features and use Mahalanobis distance at test time to see if the sample is OoD or not. Tack et al., \citep{tack2020csi} interestingly proposes that certain augmentations of the same image could be treated as OoD samples. These augmentations are supposed to be borderline cases of OoD samples and making it different from original images should lead to more robust features for the detection of OoD examples. They also used SIMCLR for contrastive learning.  
In comparison to the above-mentioned methods, the proposed approach is generic and could be applied to different diseases (such as malaria and skin cancer),  does not require OoD samples during training, introduce a new way of OoD ranking, and have good OoD classification results as compared to several competitive baselines.

\begin{figure*}
\includegraphics[page={4},width = \linewidth]{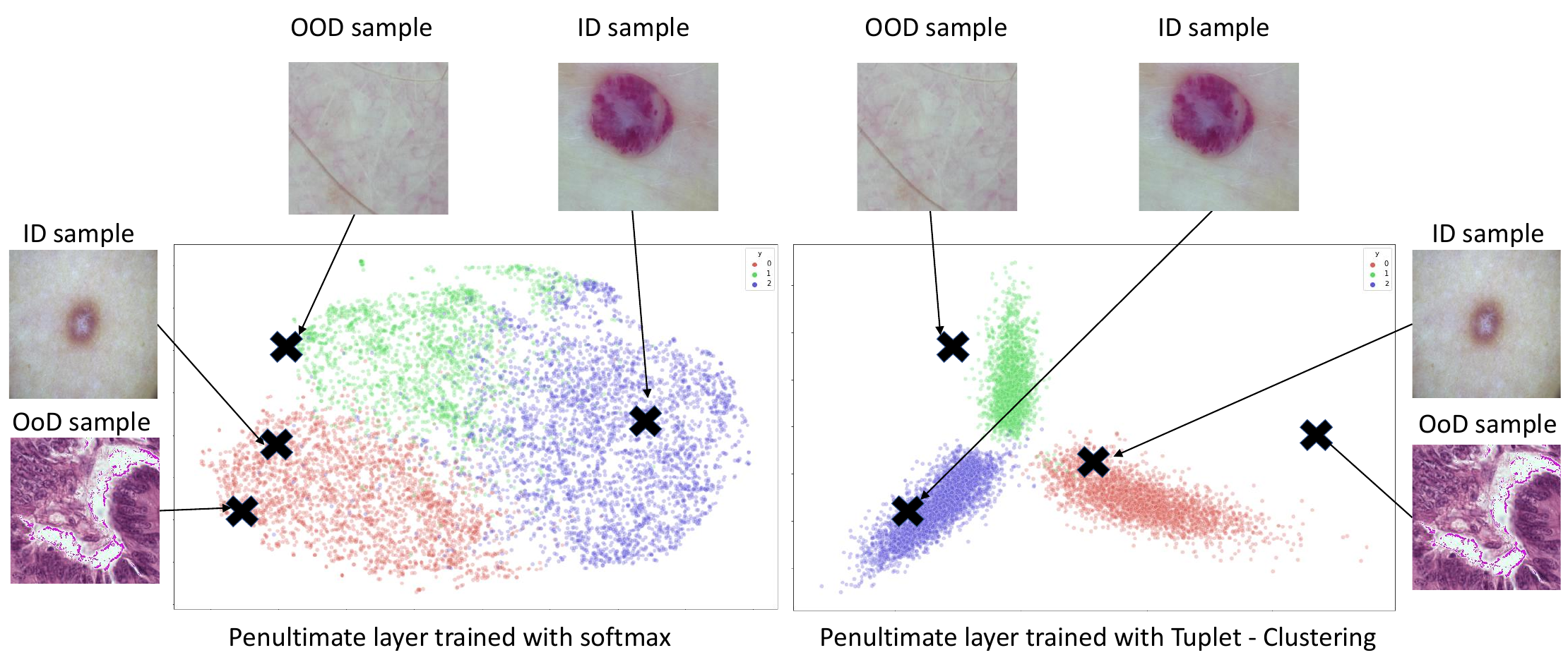}
    \caption{Left: t-SNE of penultimate layer when network is trained with standard cross entropy loss only. Right: t-SNE after tuplet loss is introduced. It can be seen that tuplet loss is helpful in demarcating OoD samples from ID samples.}
    \label{fig:clustering}
\end{figure*} 
 \begin{figure*}
    \includegraphics[page={1},width = \linewidth]{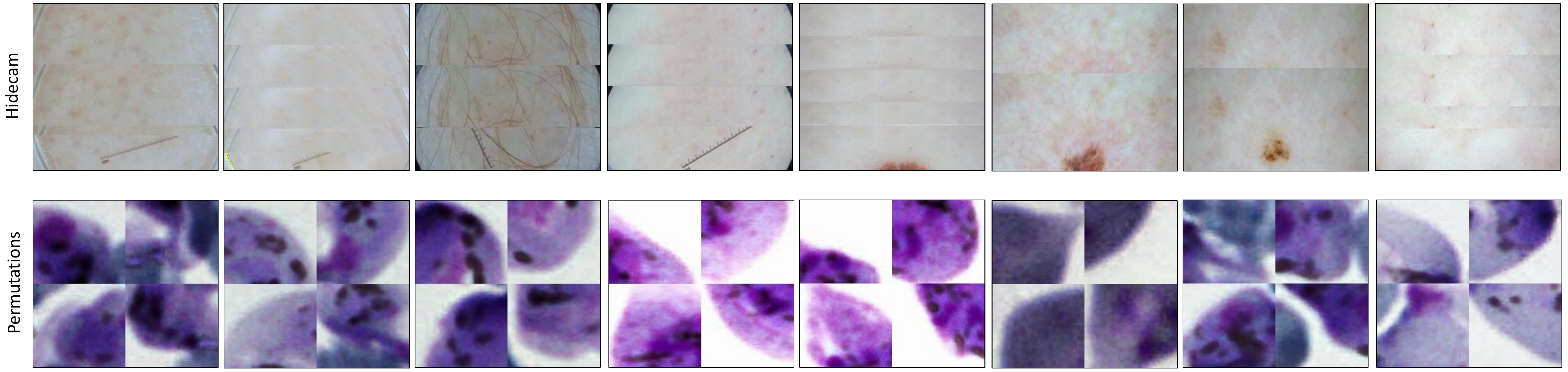}
    \caption{This figure shows OoD surrogate images generated for malaria and skin cancer. First row shows the OoD surrogates generated through HideCam for skin cancer images and second row show OoD surrogates generated through mage-Parts-Permutation for malaria images.} 
    \label{fig:surrogate}
\end{figure*}
\section{Methodology}
 
Our proposed approach to tackle the challenging problem of OoD detection is based on three observations: (1) the proposed approach should make the compact clusters of ID samples while preserving each class discriminability, (2) due to the unavailability of OoD samples during training, we need to devise some mechanism to generate OoD surrogates to train a discriminative OoD detector, (3) finally, due to limitation of K-nearest neighbors for large datasets, we should have employ better distance metric to rank OoD samples. Below, we provide details of each step of our approach.

We propose to employ class-aware metric learning such that the samples belonging to the same class are mapped closer to each other than the samples from the other class. This minimization of intra-class variance and maximization of inter-class variance is a desired property of many clustering algorithms as well. For a pre-trained network on skin cancer  {or malaria} dataset, we finetune it using {tuplet margin loss} \cite{yu2019deep}.
Tuplet loss can be seen as an extension to triplet loss where a tuplet contains multiple negative samples with the anchor and positive. The negatives are then weighted according to a scale factor. In our context, the loss for ID samples is defined as:
\begin{equation}
\label{eq:lTupInD}
    \mathcal{L}_{tup}^{ID}(a,p,n_{1},...,n_{k-1}) = \log\bigg(1+ \sum_{i=1}^{C-1} e^{s(cos\theta_{an_{i}} - cos\theta_{ap})}\bigg),
\end{equation}
where $s$ is the scaling factor for weights, $C$ is the number of classes, $a$, $p$ represents anchor and positive samples from the same   class, whereas $n_{1},...,n_{C-1}$ are samples from the other classes. Finally, $\theta_{an_{i}}$ and $\theta_{ap}$
is the angle between the samples $a$ and $p$ and samples $a$ and $n_i$ respectively. 

Minimizing the loss ${L}_{tup}^{ID}$ (Eq. \ref{eq:lTupInD}) results in compactness of space over which samples belonging to one class are mapped while increasing the inter-class separability. It can be seen in Figure  \ref{fig:clustering} that features obtained through training a network using tuplet loss are more useful for OoD detection as compared to the features trained using only cross-entropy loss.  Since the loss function is only looking at the ID samples, the mapping can still result in ID and OoD samples being near to each other in latent space.

\subsection{ Generating OoD surrogates}

 {${L}_{tup}^{ID}$ (Eq. \ref{eq:lTupInD}) clusters the in-distribution (ID) samples, however it does not contain any knowledge about out-of-distribution (OoD) samples. 
 Therefore, to mitigate the shortcoming of the ${L}_{tup}^{ID}$ (Eq. \ref{eq:lTupInD}), we propose to generate OoD samples with the help of existing ID samples. These surrogates will enable us to learn a boundary between ID and OoD data. 
  
 OoD sample space is extremely large, consisting of all the possible distributions that are not part of ID. Therefore instead of creating a dataset that could capture all these possible variations, we concentrate on generating OoD examples by manipulating the ID samples to capture the boundary of the ID manifold.
  
 We generate OoD surrogates in two ways i.e., HideCam for skin cancer and Image-parts-Permutation  {\cite{tack2020csi}}   for malaria. 
 Skin cancer images contain dense confined regions representing cancerous class, therefore, our proposed method HideCam removes class salient regions to make OoD. On the other hand, malaria images do not contain any such dense pattern, however, the overall structural information is important. Therefore, we employ image part-based permutation to generate OoD for malaria images. Below, we provide the details of both surrogate generation techniques.}

\begin{enumerate}[label=(\Alph*)]

\item \noindent\textbf{HideCam: } We generate the class-dependent surrogate images. Inspired from the observations of \cite{hendrycks2018deep} and \cite{ren2019likelihood}, we create  surrogate OoD samples such that non-salient information is preserved and class representative information is removed.

We used gradient-based class activation maps (CAM) \cite{selvaraju2017grad} from the pre-trained network, to identify the regions that are class salient. Gradient-based CAM highlights the region of importance pertaining to the specific class. Once we have the information about the spatial position of cancer (or any other class for that matter), we crop the healthy region from the same image and overwrite the cancerous regions with these healthy crops. The resulted images are used as OoD surrogates for skin cancer. 


\item \noindent\textbf{Image-Parts-Permutation}: Since certain transformations/augmentations of the image can lead to good representatives of OoD data \cite{tack2020csi}, we generate new images through image-parts based permutations. Specifically, to generate an OoD image for malaria, the image is simply divided into \textit{n} equal regions, and then those regions are randomly shuffled. Intuitively, this shifts the input distribution. 
\end{enumerate}

The samples for HideCam and Image-Parts-Permutation are shown in Figure \ref{fig:surrogate}. Similar to Eq \ref{eq:lTupInD}, tuplet loss is applied using ID samples and OoD surrogates to bring ID samples close to each other and OoD samples far from each other in the feature space.
\begin{figure}
    \centering
        
    \includegraphics[page={2},scale = 0.23]{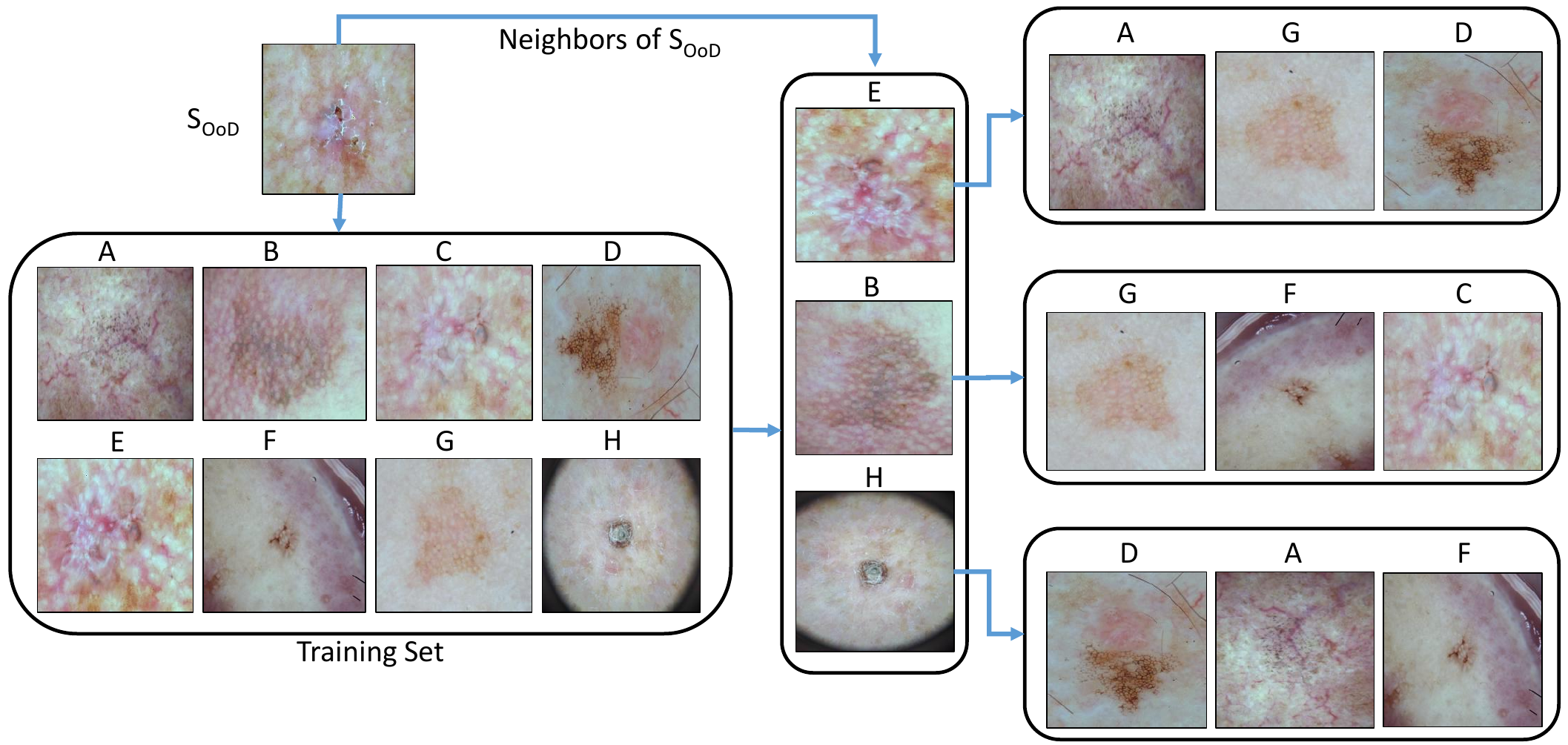}
    \caption{This figure illustrates the K-reciprocal nearest neighborhood of an OoD image for the skin cancer dataset. For $S_{OoD}$ image, we show three nearest neighbors from the training set. After that, we show three neighbors of each neighbor of $S_{OoD}$. It can be seen that $S_{OoD}$ does not appear in anyone's neighbor list, hence marked as OoD sample.} 
    \label{fig:knn_images}
\end{figure}
\begin{figure}
    \centering
        
    \includegraphics[page={5},scale = 0.35]{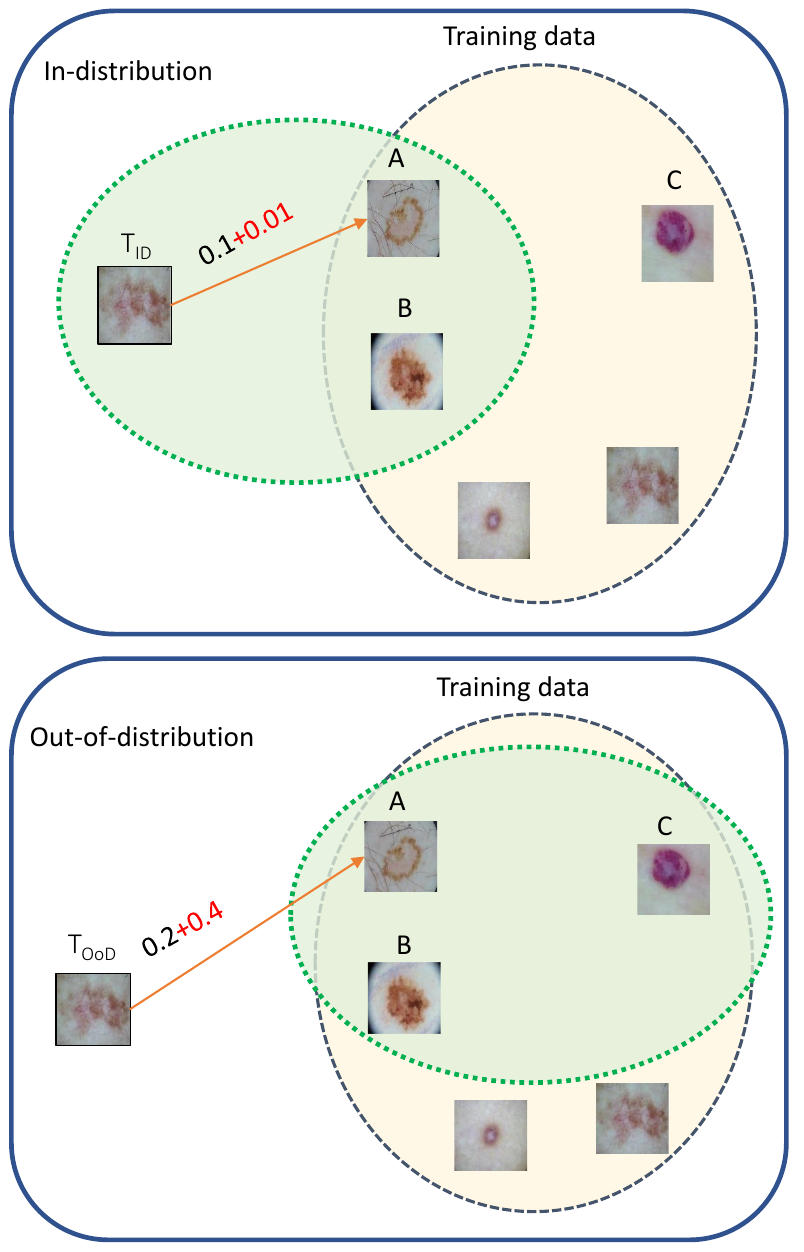}
    \caption{Intuition behind k-reciprocal neighbors score for OoD detection. The red-arrow indicates the nearest neighbor of input sample. Green color shows the nearest neighbors of training example `A'. Distance in red color shows the distance  {added} because of k-reciprocal neighbors. (a) Since $x_{ID}$ is in the neighborhood of training example `A', a smaller distance is added. (b) $x_{OoD}$, on the contrary, is not in the neighborhood of A and hence a higher value of distance is added. 
    }
    \label{fig:k_res}
\end{figure}

\subsection{Objective function}
   
Given a network trained for classification task, we fine-tune last few layers using the following loss function:
\begin{equation} 
\begin{aligned}
\sum_{a,p,n_{i}}  \mathcal{L}_{tup}^{ID}(a,p,n_{1},..,n_{C-1}) &+ \\
\sum_{a,p,n_{s_{i}}} \mathcal{L}_{tup}^{OoD}(a,p,n_{s_1},..,n_{s_{C-1}}) &+ 
\sum_{a} \mathcal{L_{CE}} (g(a),y).
\end{aligned}
\label{eq:obj}
\end{equation}
 
Similar to Eq. \ref{eq:lTupInD}  $a,p,n$ represent anchor, positive and negative of the tuplet respectively.
$n_{s_i}$ represents OoD surrogates created from the method described in section 3.1. $\mathcal{L_{CE}}$ is cross-entropy loss. $\mathcal{L}_{tup}^{ID}$ and $\mathcal{L}_{tup}^{OoD}$ represents tuplet loss with ID only and with ID + OoD surrogates respectively. In this equation, the first term is a tuplet loss where positives and negatives come from the same and different classes with respect to the anchor. In the second term,  generated surrogate OoD samples constitute negatives in the tuplet. This term is responsible for the separation of OoD surrogates from ID samples. This second term is the most important one since clustering classwise samples alongside distancing OoD surrogates is what makes the identification of OoD easier. The third term ensures that, in addition to OoD detection, the classifier learns to discriminate between different classes.

\begin{figure*}
    \centering

    \includegraphics[page={2},scale = 0.32]{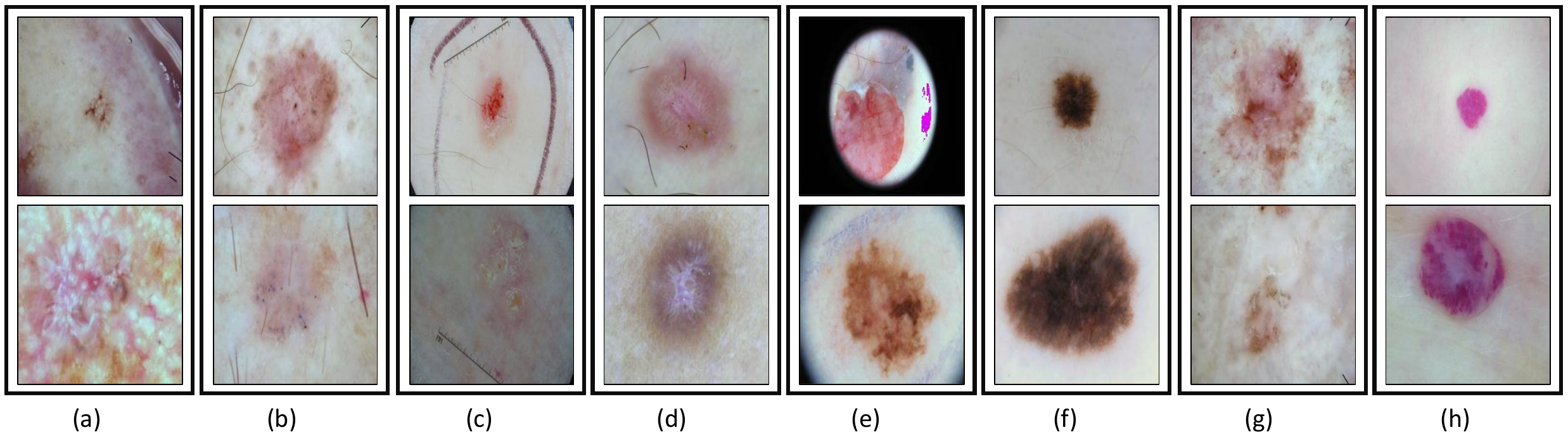}
    \caption{ This figure shows the images of different types of skin cancer disease. In figure (a) stands for Actinic keratosis (AK), (b) for Basal  cell carcinoma (BCC), (c) for Benign  keratosis (BKL), (d) for Dermatofibroma (DF), (e) for Melanoma (MEL), (f) for Melanocytic nevus(NV), (g) for Squamouscell  carcinoma (SCC) and (h) for Vascular lesion Squamous  cellcarcinoma (VASC).    } 
    \label{fig:skin}
\end{figure*}

\begin{figure*}
        
    \includegraphics[page={3},width = \linewidth]{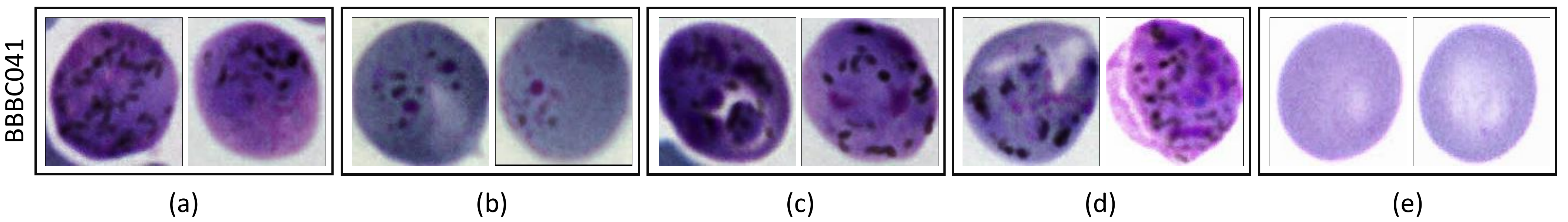}

    \caption{This figure shows images of different types of malaria-infected cells and healthy red blood cells for the BBBC041 dataset, where (a) stands for Gametocyte, (b) for Ring, (c) for Schizont, (d) for Trophozoit, and (e) for Healthy red blood cells.} 
    \label{fig:imlchl}
\end{figure*}
\subsection{Inference - OoD score}
Instead of relying on the output of the softmax layer to classify the OoD, we use the consistency in the retrieval result as the parameter to identify OoD. {A simple OoD score for an input sample could be obtained by finding its nearest neighbors in the training set, and deciding if the sample and nearest neighbors have a distance less than some specific threshold or not.}
Due to the sensitivity of naive nearest neighbors scoring to outliers and lower accuracy when dealing with high dimensional large dataset  \cite{JBI-KNN,balsubramani2019adaptive}, we propose to use distance based on K-reciprocal neighbors \cite{qin2011hello}, employing a combination of Jaccard
and Euclidean distance \cite{zhong2017re}.


To make our paper self-contained, below we describe K-reciprocal neighbors in detail.
 Suppose, we represent K-nearest neighbors of the sample $x$ by $KN(x,k)$, then K-reciprocal neighbors of $x$ can be written as:
 \begin{equation}
    R(x,K) = \{y_i \mid (y_i \in KN(x,K)) \wedge (x \in KN(y_i,K))\}.
    \label{k_rec_eq}
\end{equation}
where $y_i$ is an image from training set which is neighbor of query image $x$ and query image also appears in the neighbor's list of $y_i$.
It can be easily seen that K-reciprocal neighbors is more stringent than naive nearest neighbors in the sense that mutual neighborhood is required for being a K-reciprocal neighbors.
Furthermore, the Jaccard distance between two samples $x$ and $tr$ can now be defined as: 
\begin{equation}
    \mathcal{J}(x,tr) = 1 - \frac{|R^*(x,k) \cap R^*(tr,k)|}{|R^*(x,k) \cup R^*(tr,k)|},
\end{equation}
where $\mathcal{J}$ stands for Jaccard distance and where $R^*(x,k)$ represents the locally expanded neighborhood created from $R(x,k)$ \cite{zhong2017re}. The local query expansion is the process of expanding the neighborhood of a sample beyond the naive nearest neighbors. It does so by taking the nearest neighbors of query sample and taking the union of the original neighbors of query and neighbors of the neighbors.
It can be observed that for the samples having mutual K-reciprocal neighbors, $\mathcal{J}$ will be smaller. Therefore, for two samples to have a lower distance, it must be the case that their neighborhoods are mutual which is hard to achieve if the sample is far away from the training sample.

Finally, we compute the distance of testing sample $x$ with training sample $tr$ is computed using using the following equation:
\begin{equation}
    d^*(x,tr) = (1-\lambda)\mathcal{J}(x,tr) + \lambda d(x,tr),
    \label{eq:dist}
\end{equation}
where $d^*$ is the weighted summation of $\mathcal{J}$(Jaccard distance) and $d$ (Euclidean distance). $\mathcal{J}$ is higher when $x$ is \textbf{not} in the neighborhood of $tr$. Intuitively, it will enlarge the distance of sample which is far away from training sample $tr$. The demonstration of this idea is depicted in Figure \ref{fig:knn_images} and Figure \ref{fig:k_res}.

 For robustness, the OoD score of a sample $x$, $OoD_s$(x), is computed by taking the median distance of nearest neighbors i.e.,  
\begin{equation}
    OoD_s(x) = median(d^*(x,tr_1),d^*(x,tr_2),...,d^*(x,tr_{n}))
    \label{eq:median}
\end{equation}
where $tr_i$ shows the $i^{th}$ closest neighbor of $x$ based on $d^*$ and $n$ is set to 15 in experiments.

\section{Experiments}
The goal of our experiments is to thoroughly evaluate the proposed approach on skin and malaria OoD datasets over the different evaluation metrics and analyze different components of our approach.
 
\subsection{Datasets}
 \subsubsection{ISIC 2019 Dataset}
We evaluate our approach using the setup proposed by  \cite{pacheco2020out}, where two standard skin cancer datasets, ISIC 2018 and ISIC 2019 \cite{codella2019skin,tschandl2018ham10000,codella2018skin}  are used as ID data. ISIC 2019 dataset contains 25,331 images of eight skin cancer classes i.e., Melanoma (MEL), Melanocytic nevus (NV), Basal cell carcinoma (BCC), Actinic keratosis (AK), Benign keratosis (BKL),  Dermatofibroma (DF), Squamous cell carcinoma (SCC) and Vascular lesion Squamous cell carcinoma (VASC). The samples of each of them are shown in Figure \ref{fig:skin}.

We use six different OoD datasets \cite{pacheco2020out} for the model trained on skin images of ISIC. The details of each of them are as follows: 
\begin{itemize}
\item \textbf{Imagenet} contains 3000 randomly selected images from orignal ImageNet dataset.
\item \textbf{NCT } contains  1350 colorectal cancer images randomly selected from NCT-CRC-HE-7K \cite{kather2019predicting}.
\item \textbf{BBOX } contains 2025 ISIC 2019 images where the lesion is covered by a black bounding box.
\item \textbf{BBOX 70 }is the same as BBOX except that this dataset has atleast 70\% of the lesion covered by a black bounding box. 
\item \textbf{Derm-skin } represents dermoscopic 1,565 healthy skin  images which are taken from ISIC 2019. 
\item \textbf{Clinical} comprises 723 clinical healthy skin images. 
\end{itemize}

\subsubsection{BBBC041 Malaria Dataset}
 {This dataset is taken from Broad Bioimage Benchmark Collection \cite{ljosa2012annotated} and contains 1328 images with almost 80,000 cells.
Blood smears were stained with Giemsa reagent and images were captured using a microscopic camera. Finally, the infected cells were annotated by a team of experts. The data consists of two classes of healthy cells (i.e. red blood cells and leukocytes) and four classes of malaria-infected cells i.e. gametocytes (125 images), rings (418 images), trophozoites (1270 images), and schizonts (1270 images). The data has a heavy imbalance towards healthy red blood cells as compared to healthy leukocytes and malaria-infected cells where healthy cells making up over 95\%   of all cells. Some samples of this are shown in the second row of Figure \ref{fig:imlchl}.  Similar to \cite{pacheco2020out}, we introduce the following OoD datasets for malaria. The details of each of them are as follows:} 

\begin{itemize}

\item \textbf{Imagenet} contains 3000 randomly selected images from original ImageNet dataset.
\item \textbf{NCT } contains  1350 colorectal cancer images randomly selected from NCT-CRC-HE-7K \cite{kather2019predicting}.
\item \textbf{Coin fusion} contains 500 blended images of coin and healthy red blood cells. A coin and cell are fused using the Laplacian of Gaussian pyramids.
\item \textbf{Flag fusion} contains 192 blended images of flags (of different countries) with healthy red blood cells. The process of creating these was similar to coin fusion.
\item \textbf{Healthy red blood cell} contains 3868 healthy red blood samples from the BBBC041 dataset. This is the most difficult case of all.

\end{itemize}
\begin{figure}[h]
    \includegraphics[page={1}, width = \linewidth]{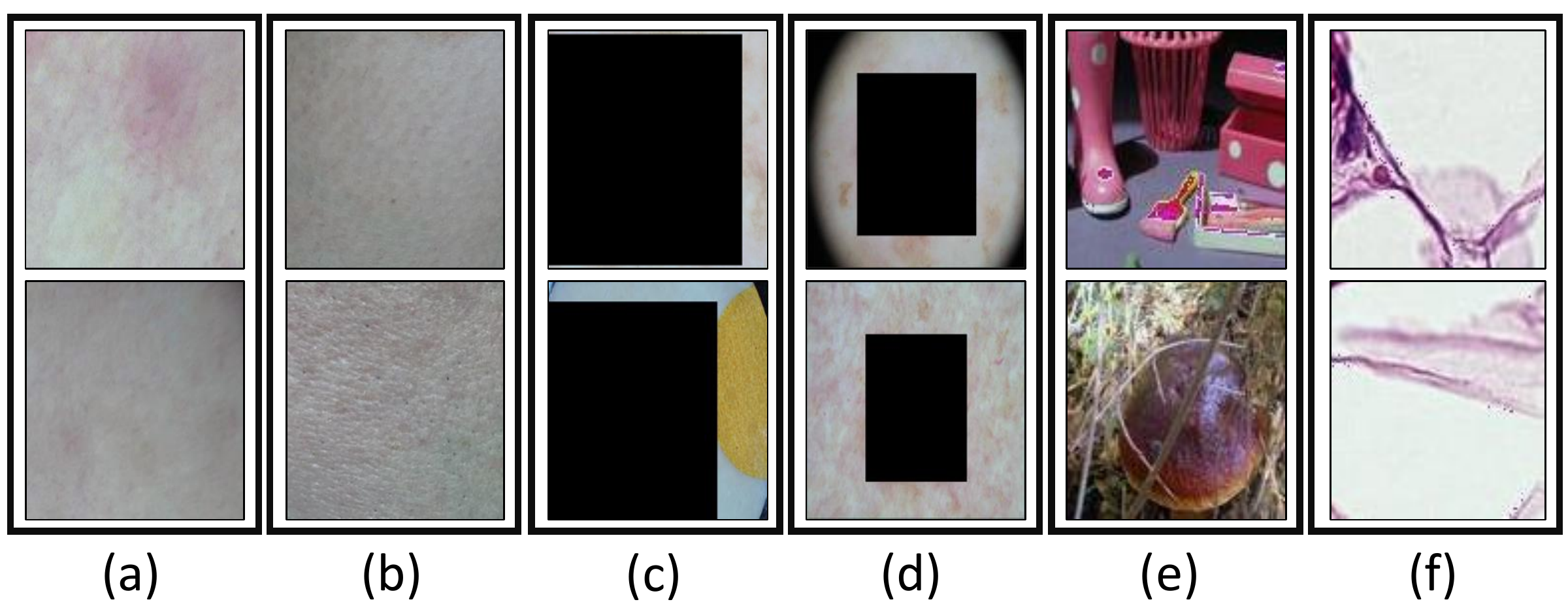}
    \caption{This figure shows OoD images generated for skin dataset 
    . Where (a) stands for Derm-skin, (b) for Clinical, (c) for BBOX70, (d) for BBOX and (e) for Imagenet and (f) for NCT.
    } 
    \label{fig:ood_skin}
\end{figure}
\begin{figure}[h]
    \includegraphics[page={1}, width = \linewidth]{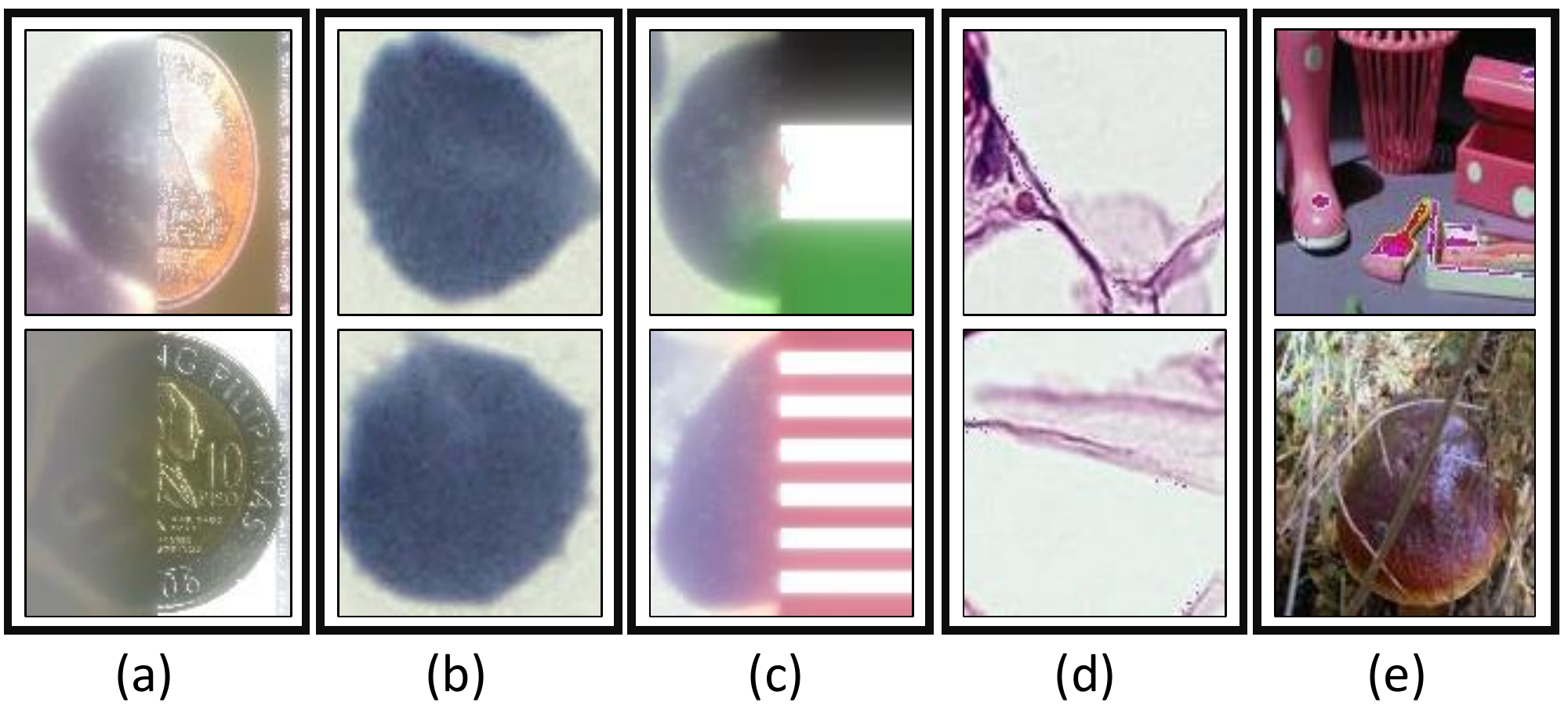}
    \caption{Figure shows OoD images generated for malaria datasets. Where (a) stands for Coin fusion, (b) for Healthy red blood cells, (c) for Flag fusion, (d) for NCT and (e) for Imagenet.} 
    \label{fig:malaria_ood}
\end{figure}

The samples of these different OoD samples for skin and malaria are shown in Figure \ref{fig:ood_skin} and Figure \ref{fig:malaria_ood} respectively. In addition to evaluating the model on the aforementioned OoD datasets, we also explored the efficacy of the approach by leaving one class at a time as an OoD class. This means that we train the network on seven (for example in the skin dataset) classes and consider the eighth as an OoD class.

\begin{table*}[t]
\renewcommand{\arraystretch}{1.6}
\resizebox{\textwidth}{!}{%
\begin{tabular}{|c|c|cccccc|} 
\hline
 \textbf{Dataset}  & \textbf{Metric}  & \textbf{Baseline[US] }  & \textbf{ODIN[S] }  & \textbf{Mahalanobis[S] }  & \textbf{Gram-OoD[US] }  & \textbf{Gram-OoD*[US] }  
 & \textbf{Ours [US]} \\
\hline
 \multirow{3}{*}{\large Derm-skin} & \textbf{TNR @ TPR 95\%}  &  \large 22.8  & \large 46.2 & \large 81.47 & \large 78.0  & \large 76.1 
 & \large \textbf{95.59}  \\[3pt]
 & \textbf{AUROC}  & \large 74.4 & \large 86.8 & \large 96.2 & \large 96.5  & \large 95.8 
 & \large \textbf{98.66}  \\[3pt]
 & \textbf{Detection acc}  & \large 67.3 & \large 78.3 & \large 89.7 & \large 90.9  & \large 89.3 
 & \large \textbf{96.50}  \\[3pt]
\hline 
\multirow{3}{*}{\large Clinical} & \textbf{TNR @ TPR 95\%}  & \large 18.5 & \large 25.2 & \large 81.7 & \large 82.8 & \large 83.1 
& \large \uline{85.33}  \\[3pt]
 & \textbf{AUROC}  & \large 72.5 & \large 69.5 & \large 96.1  & \large \uline{96.6} & \large \uline{96.6} 
 & \large 95.59 \\[3pt]
 & \textbf{Detection acc}  & \large 67.3 & \large 65.8 & \large 90.8 & \large \uline{91.1} & \large 90.9 
 & \large 90.27 \\[3pt] 
\hline
\multirow{3}{*}{\large Imagenet} & \textbf{TNR @ TPR 95\%}  & \large 9.30 & \large 50.0 & \large \textbf{99.9} & \large 80.7 & \large 88.4 
& \large \uline{90.86}  \\[3pt]
 & \textbf{AUROC}  & \large 59.1 & \large 83.8 & \large \textbf{99.9}  & \large 97.0 & \large \uline{97.7} 
 & \large 97.17 \\[3pt]
     & \textbf{Detection acc}  & \large 56.6 & \large 78.1 & \large \textbf{99.1} & \large 92.0 & \large \uline{97.9}
     & \large 94.15 \\[3pt] 
\hline
\multirow{3}{*}{\large BBOX} & \textbf{TNR @ TPR 95\%}  & \large 27.9 & \large 68.8 & \large \uline{94.8}  & \large 88.0 & \large 88.1 
& \large \textbf{97.60}  \\[3pt]
 & \textbf{AUROC}  & \large 77.3 & \large 90.6 & \large \uline{98.3} & \large 98.1 & \large 97.5
 & \large \textbf{98.89} \\[3pt]
 & \textbf{Detection acc}  & \large 69.8 & \large 83.7 & \large \uline{95.3} & \large 94.5 & \large 94.0
 & \large \textbf{98.33} \\[3pt] 
\hline
\multirow{3}{*}{\large BBOX-70} & \textbf{TNR @ TPR 95\%}  & \large 36.6 & \large 99.3 & \large \textbf{100} & \large \uline{99.9}  & \large \textbf{100} 
& \large \textbf{100} \\[3pt]
 & \textbf{AUROC}  & \large 89.4 & \large 99.8 & \large \textbf{100} & \large 99.7 & \large \uline{99.9} 
 & \large \textbf{100}  \\[3pt]
 & \textbf{Detection acc}  & \large 84.9 & \large 98.1 & \large \uline{99.9} & \large 99.0 &\large 100 
 & \large \textbf{100}  \\[3pt] 
\hline
\multirow{3}{*}{\large NCT} & \textbf{TNR @ TPR 95\%}  & \large 1.44 & \large 32.5 & \large 98.7 & \large 98.9 & \large \textbf{99.9} 
& \large \uline{99.30} \\[3pt]
 & \textbf{AUROC}  & \large 36.7 & \large 82.0 & \large 98.9 & \large 99.4 & \large \uline{99.7} 
 & \large \textbf{99.85} \\[3pt]
 & \textbf{Detection acc}  & \large 50.1 & \large 75.0 & \large \textbf{98.7} & \large 97.1 & \large \uline{98.5} 
 & \large 92.10 \\[3pt] 
\hline
\multirow{3}{*}{\large Average (S)} & \textbf{TNR @ TPR 95\%}  & - & \large 53.67 & \large \uline{92.75}  & - & - 
& \large \textbf{94.79}  \\[3pt]
 & \textbf{AUROC}  & - & \large 85.42 & \large \uline{98.23}  & - & - 
 &\large \textbf{98.36}  \\[3pt]
 & \textbf{Detection acc}  & - & \large 79.83 & \large \uline{95.47}  & - & - 
 & \large \textbf{96.41}  \\[3pt] 
\hline
\multirow{3}{*}{\large Average (US)} & \textbf{TNR @ TPR 95\%}  & \large 19.42 & - & - & \large 88 & \large \uline{89.27}  
& \large \textbf{94.79}  \\[3pt]
 & \textbf{AUROC}  & \large 68.23 & - & - & \large \uline{97.9}  & \large 97.87 
 &\large \textbf{98.36}  \\[3pt]
 & \textbf{Detection acc}  & \large 66.0 & - & - & \large 94.1 & \large \uline{94.97}  
 & \large \textbf{96.41}  \\[3pt]
\hline
\end{tabular}
}

\caption{Comparison with state-of-the-art OoD detection algorithms for ISIC 2019 dataset. \textbf{US} and \textbf{S} represents unsupervised and supervised algorithms. 
Our algorithm being unsupervised does not require any real OoD sample. Best Scores are shown in bold, second best is underlined.}
\label{isic_2019}
\end{table*}

\begin{table}[b]
\centering
\resizebox{\columnwidth}{!}{%
 \begin{tabular}{|c | c | c | c |}
\hline
\textbf{OoD Class} & TNR @ TPR 95\% & AUROC & Detection accuracy \\
\hline
BKL & \small 20.60 & \small 67.29 & \small 63.90 \\
AK & 14.30 & 62.40 & 62.35 \\
BCC & 36.14 & 74.24 & 69.90\\
DF & 12.61 & 65.59 & 63.25\\
MEL & 12.40 & 66.25 & 62.67\\
NV & 7.36 & 70.14 & 67.21\\
SCC & 17.43 & 61.80 & 61.84\\
VASC & 9.74 & 61.97 & 59.53\\
\hline
 \end{tabular}}
\caption{Results when one class is kept as OoD and rest are kept as ID.  }
 \label{isic_one_vs_rest}
 \end{table}
 
\begin{table}[b]
\resizebox{\columnwidth}{!}{%
 \begin{tabular}{|c | c | c | c |}
\hline
\textbf{OoD dataset} & TNR @ TPR 95\% & AUROC & Detection accuracy \\
\hline
Derm-skin & \small 100 & \small 100 & \small 100\\
Clinical & 100 & 99.90 & 99.90\\
Imagenet & 99.30 & 99.70 & 99.30\\
BBOX & 99.80 & 99.80 & 99.70\\
BBOX-70 & 100 & 100 & 100\\
NCT & 99.60 & 99.90 & 99.60\\
\hline
Average &  99.78 & 99.88 & 99.75\\
\hline
Standard dev & 0.0029 & 0.0012 & 0.0027\\
\hline

 \end{tabular}}
 \caption{Results on OoD data when ID data is from ISIC 2018}
 \label{isic_2018}
 \end{table}

\subsubsection{Evaluation Metrics}
To measure the efficiency of our approach to differentiate between ID and OoD examples, we use three evaluation metrics which are described below for completeness.
\begin{itemize}
  \item \textbf{TNR@95TPR} is the likelihood that an OoD sample is accurately identified with the true positive rate (TPR) as high as 95\%. The true positive rate can be computed as TPR = TP/(TP + FN), where TP and FN represent true positive and false negative respectively \cite{sastry2019detecting}. 
  \item \textbf{Detection Accuracy} measures the maximum achievable classification accuracy across all possible thresholds in distinguishing among ID and OoD samples \cite{sastry2019detecting}.
  \item \textbf{AUROC} is the measure of the area under the curve of true positive rate vs false positive rate \cite{sastry2019detecting}.
\end{itemize}

\subsection{Implementation details}
In this section, we provide implementation details of our approach. The proposed approach is divided into two stages. In the first phase, we train DenseNet-121 \cite{huang2018densely}  on the ID data (ISIC and malaria) for classification purposes. For both datasets, we use the weighted cross-entropy loss function to account for class imbalance i.e., minority classes were given higher weight depending upon the frequency. The model is trained for around 100 epochs with early stopping. As an optimizer,  stochastic gradient descent is employed with a learning rate of 0.001 and weight decay of 0.001. The above-mentioned settings were the same for both skin and malaria datasets. Due to the large class imbalance in the malaria dataset, we augment the minority class (gametocytes) samples with rotations of different angles. In the second phase, we took the network trained in the first phase and trained it further with the loss function given by Equation \ref{eq:obj}. We used HideCam to generate OoD surrogates for skin cancer and Image-Parts-Permutation for malaria-related experiments. We finetuned the network in this phase for around 200 epochs. We used the default margin of 5.73 for tuplet loss and use a learning rate of $1e^-3$. \newline

For inference, we compute the distance of the test sample with every training sample using  Equation \ref{eq:dist}. As suggested in \cite{zhong2017re}, for K-reciprocal, we use the 15 nearest neighbors to be considered for a query and 6 nearest neighbors for the process of local query expansion. Finally, we use $\lambda$ = 0.3. These values were kept the same for both skin and malaria-related experiments which shows that the algorithm is not heavily dependent on finetuning these hyper-parameters.

In addition to testing on the OoD dataset introduced in section 4.1, we took 25\% of samples from original ISIC and BBBC401 datasets as testing ID samples. Following \cite{pacheco2020out}, TNR @ TPR 95\%, AUROC, $\&$ Detection accuracy are used as evaluation metrics.

\begin{table*}[t]
\renewcommand{\arraystretch}{1.3}
\centering
\resizebox{\linewidth}{!}{%
\begin{tabular}{|c|c|cccccc|} 

\hline
 \textbf{Dataset}  & \textbf{Metric}  & \textbf{Baseline[US] } & \textbf{ODIN[S] }  & \textbf{Mahalanobis[S] } &\textbf{Gram-OOD[US] } & \textbf{Gram-OOD*[US] }  
 & \textbf{Ours[US]} \\
\hline

 \multirow{3}{*}{ RBC Healthy} & \textbf{TNR @ TPR 95\%}  & \small 2.68 & \normalsize 23.97 & \normalsize \uline{49.35}  & \normalsize 58.59 & \normalsize 41.9 
 & \normalsize \textbf{78.30}   \\[3pt]
& \textbf{AUROC}  & \normalsize 48.95 &  \normalsize 78.71 & \normalsize 90.05  & \normalsize 82.64 & \normalsize \uline{91.8} 
& \normalsize \textbf{94.05}  \\[3pt]
& \textbf{Detection acc}  & \normalsize 53.18 &  \normalsize 73.45 & \normalsize 83.17 & \normalsize 78.18 & \normalsize \uline{86.61} 
& \normalsize \textbf{90.19}  \\[3pt]
\hline

 \multirow{3}{*}{\normalsize Coin fusion} & \textbf{TNR @ TPR 95\%}  &  \normalsize 29.06 & \normalsize 83.33 & \normalsize \textbf{100} & \normalsize 95.49 & \normalsize \uline{98.81} 
 & \normalsize 95.20   \\[3pt]
 & \textbf{AUROC}  & \normalsize 87.8 &  \normalsize 96.09 & \normalsize \textbf{100} &\normalsize 98.72 &  \normalsize \uline{98.67} 
 &  \normalsize 96.86  \\[3pt]
 & \textbf{Detection acc}  & \normalsize 81.8 &  \normalsize 90.36 & \normalsize \textbf{100} & \normalsize 95.64 & \normalsize \uline{98.05} 
 & \normalsize 95.58  \\[3pt]
\hline

 \multirow{3}{*}{\normalsize Flag fusion} & \textbf{TNR @ TPR 95\%}  &  \normalsize 18.75 &  \normalsize 83.33 & \normalsize \textbf{100}  & \normalsize 97.72 &  \normalsize \uline{99.58} 
 & \normalsize 84.60   \\[3pt]
 & \textbf{AUROC}  & \normalsize 85.95 &  \normalsize 97.25 & \normalsize \textbf{100} & \normalsize 99.36 & \normalsize \uline{99.12} 
 & \normalsize 95.28  \\[3pt]
 & \textbf{Detection acc}  & \normalsize 80.51 &  \normalsize 91.51 & \normalsize \textbf{100} & \normalsize 97.02 &  \normalsize \uline{98.49} 
 & \normalsize 91.56  \\[3pt]
\hline

 \multirow{3}{*}{\normalsize Imagenet} & \textbf{TNR @ TPR 95\%}  &  \normalsize 18.44 &  \normalsize 73.82 & \normalsize \textbf{100}  & \normalsize 88.82 & \normalsize \uline{98.21} 
 & \normalsize 97.60   \\[3pt]
 & \textbf{AUROC}  & \normalsize 84.21 &  \normalsize 92.65 & \normalsize \textbf{99.97} & \normalsize 97.74 & \normalsize 98.8 
 & \normalsize \uline{99.07}  \\[3pt]
 & \textbf{Detection acc}  & \normalsize 79.3 &  \normalsize 86.88 & \normalsize \textbf{99.7} & \normalsize 92.22 & \normalsize \uline{97.28} 
 & \normalsize 96.79  \\[3pt]
\hline

 \multirow{3}{*}{\normalsize NCT} & \textbf{TNR @ TPR 95\%}  &  \normalsize 18.08 &  \normalsize 71.09 & \normalsize \textbf{100}  & \normalsize 95.94 & \normalsize 98.51 
 & \normalsize \uline{99.48}   \\[3pt]
 & \textbf{AUROC}  & \normalsize 77.78 &  \normalsize 92.98 & \normalsize \textbf{99.97} & \normalsize 99.16 & \normalsize 98.94 
 & \normalsize \uline{99.10}  \\[3pt]
 & \textbf{Detection acc}  & \normalsize 73.08 &  \normalsize 85.65 & \normalsize \textbf{99.7} & \normalsize 95.56 & \normalsize 97.44 
 & \normalsize \uline{98.020}  \\[3pt]
 
\hline
\multirow{3}{*}{ Average (S)} & \textbf{TNR @ TPR 95\%}  & - & 67.11 & \uline{89.87} & - & - 
& \textbf{91.04} \\[3pt]
 & \textbf{AUROC}  & - & 91.53 & \textbf{97.9} & - & - 
 & \uline{96.31}  \\[3pt]
 & \textbf{Detection acc}  & - & 85.57 & \textbf{96.5} & - & - 
 & \uline{94.42} \\[3pt] 
\hline
\multirow{3}{*}{Average (US)} & \textbf{TNR @ TPR 95\%}  & 17.4 & - & - & 73.03 & \uline{87.40} 
& \textbf{91.04} \\[3pt]
 & \textbf{AUROC}  & 76.93 & - & - & 92.08 & \textbf{97.47} 
 & \uline{96.31}  \\[3pt]
 & \textbf{Detection acc}  & 73.57 & - & - & 88.38 & \textbf{95.57} 
 & \uline{94.42} \\[3pt]
\hline

\end{tabular}
}
 
\caption{Comparison with state-of-the-art OoD detection algorithms for BBBC041 Malaria dataset. \textbf{US} and \textbf{S} represents unsupervised and supervised algorithms. Best Scores are shown in Bold, second best is underlined.  }
\label{kaggleoodt}
\end{table*}

\begin{table}
\centering
\resizebox{\columnwidth}{!}{%
 \begin{tabular}{|c | c | c | c |}
\hline
\textbf{OoD Class} & TNR @ TPR 95\% & AUROC & Detection accuracy \\
\hline
Ring & 5.98  & 81.44  & 78.37  \\
Schizont & 34.76 & 84.06 & 78.55 \\
Gametocyte & 36.57 & 82.37 & 75.47\\
Trophozoite & 76.25 & 91.94 & 86.27 \\
\hline
 \end{tabular}}
 \caption{Results on BBBC041 Malaria dataset when one class is kept as OoD and rest are kept as ID}
 \label{kaggle_one_vs_rest}
 \end{table}
 
\begin{figure*}
    \centering
    \includegraphics[page={2},scale=0.4]{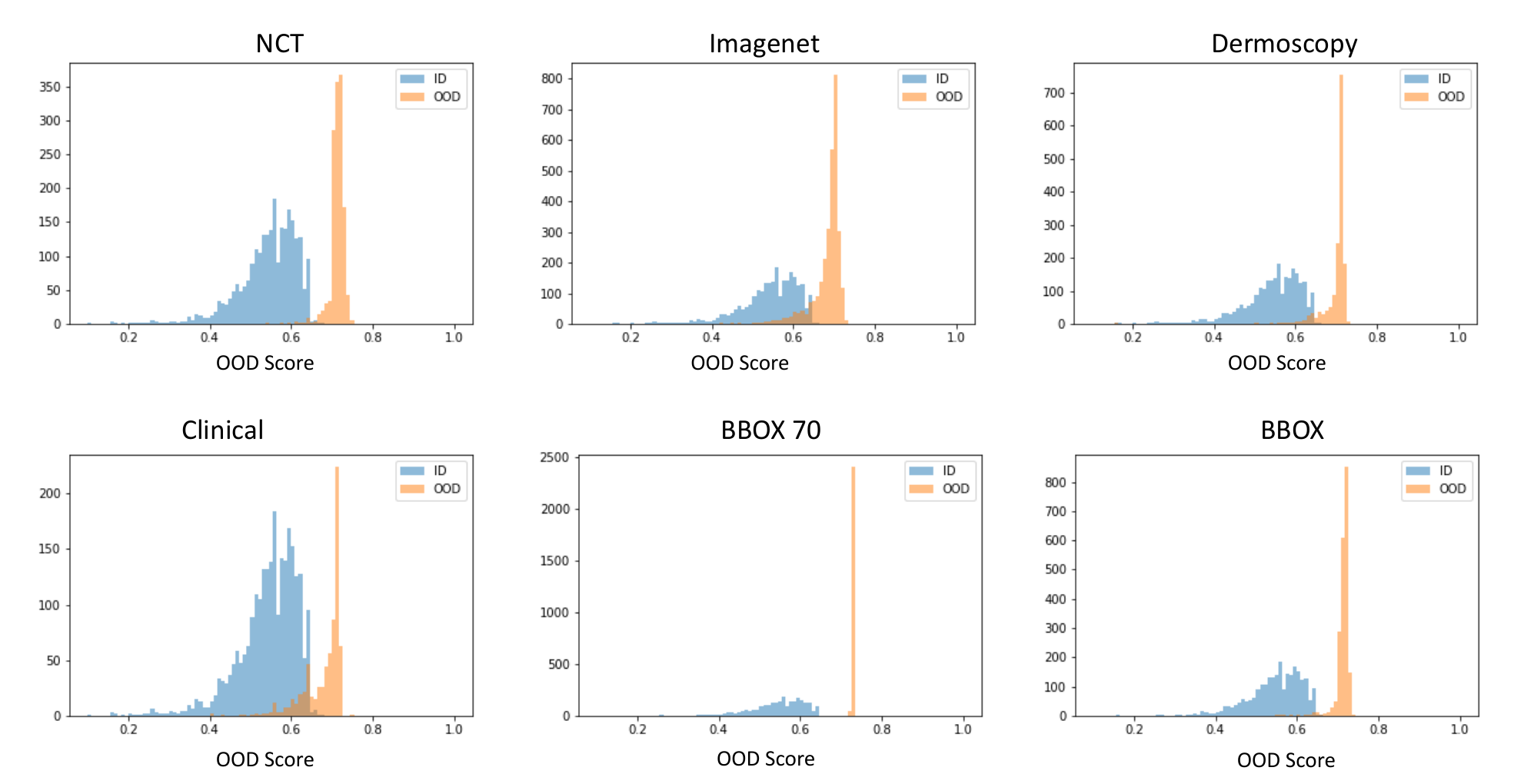}
    \caption{Histogram of OoD score for ID test set and six different OoD datasets. X-axis represents OoD scores.}
    \label{fig:hist}
\end{figure*}

\subsection{Comparison with the State-of-the-art}
We compare the proposed unsupervised OoD detection approach with supervised OoD detection approaches (ODIN \cite{liang2018enhancing} and Mahalanobis \cite{lee2018simple}) and unsupervised OoD detection approaches (Maximum softmax probability  \cite{hendrycks2016baseline}, Gram OoD \cite{sastry2020detecting} and  Gram OoD* \cite{pacheco2020out}). Note that supervised approaches assume access to labeled OoD samples during training and unsupervised approach (including ours)  assume that labeled OoD training data is unavailable. 
\subsubsection{OoD detection for Skin cancer}
{The quantitative results in Table \ref{isic_2019} demonstrate that, on all evaluation metrics, our unsupervised OoD detection approach significantly outperforms the recently published unsupervised and supervised OoD approaches. 
A clear improvement in the last column demonstrates the usefulness of our approach and enforces that clustered representation (using tuplet loss) and OoD surrogates are essential for robust OoD detection. Table \ref{isic_one_vs_rest} shows the results when one class is kept as OoD and the whole model is trained on the remaining seven classes. Since this case is extremely difficult, as the OoD (the class which is not included in the training) and ID samples look very similar, the numbers are not as good as in the case of other OoD samples as shown in Table \ref{isic_2019}.
Experimental results on the ISIC-2018 dataset are shown in Table \ref{isic_2018}. In Figure \ref{fig:hist}, we demonstrate the histogram of  OoD score (Eq.  \label{eq:median}) for ID test set (ISIC-2019) and six different OoD datasets. The histogram demonstrates the discriminative nature of the OoD score between OoD and ID samples.
}

\subsubsection{OoD detection for malaria detection}
Similar to experiments on the skin OoD dataset, we evaluate the OoD detection approach on five different malaria OoD datasets. The quantitative results of those are shown in Table \ref{kaggleoodt}. As can be seen that our approach significantly outperforms the baselines in the most difficult case, i.e. healthy red blood cells. On average, for TNR @ TPR 95\%, the proposed approach has around 4\% better results as compared to the baselines and has comparative results for AUROC and Detection accuracy. One possible reason for comparative performance on AUROC and Detection accuracy might be the fact that these metrics consider every possible threshold and there might be some thresholds where baseline models are working well. TNR metric basically works on one stringent and specific threshold which has a lot more importance and our model might be working better because we aim to work well on the case where we want good performance for detecting both in-distribution and OOD samples.

 \subsection{Ablation studies}
We analyze different components of the proposed approach to verify their effectiveness.
The experimental results in Table  \ref{ablation} on the skin ISIC 2019 dataset show that each component of our approach is important and contributes toward final accuracy.

\begin{table}[h]
\resizebox{\columnwidth}{!}{%
 \begin{tabular}{|c | c | c | c |}
\hline
\textbf{Loss} & TNR @ TPR 95\% & AUROC & Detection accuracy\\
\hline
w/o ID tuplet & 81.34 & 93.14 & 88.87\\
w/o OoD tuplet & 69.14 & 88.38 & 85.82\\
w/o CE & 79.62 & 92.95 & 89.69\\
w/o K-reciprocal & 90.07 & 96.80 & 92.60\\

Complete Approach & 94.79 & 98.36 & 96.41\\
\hline

 \end{tabular}}
 \caption{Ablation study of different components of our method. w/o means the evaluation is done without incorporating the mentioned component.}
 \label{ablation}
 \end{table}

\section{Discussion and Concluding remarks}

 {Deep neural networks have generally proven to be erroneous when it comes to detecting if a sample is coming from training distribution or not. This leads to reliability issues in deployment, especially for medical applications. We propose to tackle this by learning a representation that is not only inter-class discriminative but also has large intra-class similarities. Decreasing tuplet loss over OoD surrogates, generated from the ID dataset, helps to map out-of-distribution samples away from the class boundary. For robust OoD score estimation,  we use the median over the K-reciprocal neighbors’ distances. We have demonstrated the efficacy of the approach by obtaining state-of-the-art results on different pairs of ID and OoD datasets. We took skin cancer and malaria-contained cell images as ID and evaluated the approach on varying OoD datasets.  {We achieved state-of-the-art results, improving 5\% and 4\% in TNR$@$ TPR95\% over the previous state-of-the-art for skin cancer and malaria detection respectively.} In the future, we aim to generalize this approach to make it applicable to a wide variety of clinical datasets.}
 \newline
\noindent\textbf{Acknowledgement:}
 The project is partially supported by an unrestricted gift from Facebook, USA. The
opinions, findings, and conclusions or recommendations expressed in this publication are those of the author(s) and do
not necessarily reflect those of Facebook.

\end{document}